\documentclass[pra,twocolumn,preprintnumbers,amsmath,amssymb,nofootinbib]{revtex4}

\usepackage{amsmath,amsfonts,graphicx}
\usepackage{bbm,bm}
\usepackage{hyperref}

\newcommand{\be}{\begin{equation}}
\newcommand{\ee}{\end{equation}}
\newcommand{\bea}{\begin{eqnarray}}
\newcommand{\eea}{\end{eqnarray}}
\newcommand{\beas}{\begin{eqnarray*}}
\newcommand{\eeas}{\end{eqnarray*}}
\def\identity{{\mathbbmss 1}}

\def\la {{\langle}}
\def\ra {{\rangle}}

\def\del {\partial}

\def\half{\textstyle{1\over 2}}

\begin{document}
\def \CMP {{Commun. Math. Phys.}}
\def \PRL {{Phys. Rev. Lett.}}
\def \PL {{Phys. Lett.}}
\def \NPBProc {{Nucl. Phys. B (Proc. Suppl.)}}
\def \NP {{Nucl. Phys.}}
\def \RMP {{Rev. Mod. Phys.}}
\def \JGP {{J. Geom. Phys.}}
\def \CQG {{Class. Quant. Grav.}}
\def \MPL {{Mod. Phys. Lett.}}
\def \IJMP {{ Int. J. Mod. Phys.}}
\def \JHEP {{JHEP}}
\def \PR {{Phys. Rev.}}
\def \JMP {{J. Math. Phys.}}
\def \GRG{{Gen. Rel. Grav.}}
\preprint{CCNY-HEP-10/2}

\title{On the Casimir interaction between holes}

\author{Daniel Kabat}
\email{daniel.kabat@lehman.cuny.edu}
\author{Dimitra Karabali}%
\email{dimitra.karabali@lehman.cuny.edu}
\affiliation{Department of Physics and Astronomy,
Lehman College of the CUNY,
Bronx, NY 10468}


\author{V.P. Nair}
\email{vpn@sci.ccny.cuny.edu}
\affiliation{Physics Department,
City College of the CUNY,
New York, NY 10031}

\date{\today}

\begin{abstract}
We study the leading long-distance attractive force between two holes
in a plate arising from a scalar field with Dirichlet boundary
conditions on the plate.  We use a formalism in which the interaction
is governed by a non-local field theory which lives on the two holes.
The interaction energy is proportional to $Q_1 Q_2/r^7$ at
large separation $r$, where $Q_1$ and $Q_2$ are certain charges associated with
the holes.  We compute these charges for round and rectangular holes.
We show that the $1/r^7$ behavior is universal for separations large compared to the
linear dimensions of the holes, irrespective of the spin or interactions of the bosonic field.
We also study the interaction between two long thin slits, for which the
energy falls off as $1/r^6$.
\end{abstract}

\pacs{Valid PACS appear here}
\maketitle


\section{Introduction}

The original Casimir effect described the interaction between two parallel
conducting plates due to vacuum fluctuations of the electromagnetic field \cite{Casimir:1948dh}.
Since the pioneering work of Casimir many variants of
this effect have been studied.  For a recent review see \cite{Milton:2008st}.

In the present paper we consider a single infinite plate with two holes in
it.  We consider a scalar field with Dirichlet boundary conditions on the plate,
and study the way in which the holes modify the ground state energy of
the field.  We do this using a formalism developed in \cite{Kabat:2010nm},
where one first integrates out the scalar field in the bulk to obtain a
non-local field theory that lives in the holes.  This description makes it
easy to study the leading long-distance interaction between holes.  We find
that the force between the holes is attractive and that the interaction energy scales as $1/r^7$, where $r$ is the distance
between the holes.  The interaction energy is also proportional to the product
$Q_1 Q_2$ of certain charges associated with the holes.  These charges have units of
$({\rm length})^3$ and depend on the geometry of the holes.

An outline of this paper is as follows.  In section \ref{holes} we set up the
general formalism and extract the distance dependence.  We then compute the
charge for round and rectangular holes.  In section \ref{slits} we study the interaction
between two long thin parallel slits, using a mix of numerical and
perturbative techniques.  Section \ref{universality} gives the argument for the 
universality of the $1/r^{7}$ interaction.
We conclude with some remarks in section \ref{discussion}.

\section{Interaction between holes\label{holes}}

\begin{figure}[!b]
\begin{center}
\scalebox{.75}{\includegraphics{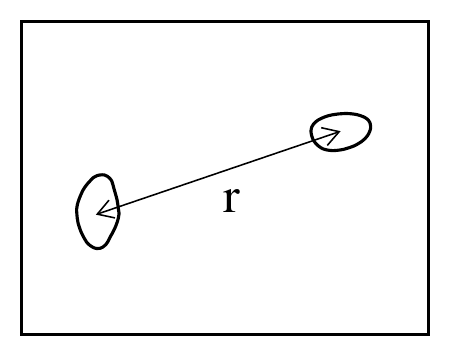}}
\end{center}
\caption{An infinite plate with two holes separated by a distance $r$.  The field vanishes at the
location of the plate.  In the holes we denote the fluctuating value of the
field by $\phi_0$.\label{plate}}
\end{figure}

Consider a scalar field $\phi$ in the presence of an infinite plate with two holes
in it, as shown in Fig.~\ref{plate}.
We first work in three Euclidean dimensions with a scalar field of mass $\mu$.
We introduce a pair of coordinates ${\bf x}$ along the plate
and a single transverse coordinate $y$. We may take the field $\phi$ to be confined to 
a large cubical box 
with the plate of interest partitioning this box into a left region and a right region.The location of the plate is at $y =0$.
We impose a Dirichlet condition on the plate.

The approach of
\cite{Kabat:2010nm} begins by integrating out the scalar field in the bulk
to write a non-local effective action for the field in the holes.  One way
to think about this effective action is to regard $y$ as Euclidean time.
The ground state wavefunctional for the field takes the standard form
\cite{Jackiw:1988sf,Hatfield:1992rz}
\be
\Psi_0[\phi] = {\rm const.} \, \exp \left(- \int d^2x \, {1 \over 2} \phi
\sqrt{-\nabla^2 + \mu^2} \, \phi\right)\,. \label{1}
\ee
Here $\phi({\bf x})$ is the value of the field on a slice of fixed $y$ and
$\nabla^2$ is the Laplacian on ${\mathbb R}^2$.  The three-dimensional
partition function of the field in the presence of the plate at $y = 0$ is then
\be
\label{Z3d}
Z_{3d} = \int {\cal D} \phi_0 \, \Psi_0^*[\phi_0] \, \Psi_0[\phi_0]
\ee
where $\phi_0$ is the value of the field at $y = 0$.  The boundary
conditions imposed by the plate are taken into account by requiring
that $\phi_0$ vanishes outside the holes.
This formalism was used in \cite{Kabat:2010nm} to obtain the diffractive edge effects for the Casimir interaction between a plate and half-plate parallel to it,
as well as for a plate and another plate perpendicular to it at a finite separation.
This gave results in agreement with numerical calculations \cite{gies} for both cases and with a special calculation for the case of perpendicular plates \cite{MIT2}.

For working out the functional integral in
(\ref{Z3d}) explicitly for the present case of two holes on the plate,
 it is useful to introduce projection operators $P_1$,
$P_2$ on the two holes.  These are operators on $L^2({\mathbb R}^2)$, defined
by
\be
\label{projection}
P_i f(x) = \left\lbrace
\begin{array}{cl}
f(x) & \hbox{\rm if $x \in \hbox{\rm hole $i$}$} \\
0 & \hbox{\rm otherwise}
\end{array}
\right.
\ee
The Gaussian functional integral in (\ref{Z3d}) then gives
\be
- \log Z_{3d} = {1 \over 2} \, {\rm Tr} \, \log \left(
\begin{array}{cc}
{\cal O}_{11} & {\cal O}_{12} \\
{\cal O}_{21} & {\cal O}_{22}
\end{array}\right) \label{Z3da}
\ee
where ${\cal O}_{ij} = P_i \sqrt{-\nabla^2 + \mu^2} P_j$.  To extend this
result to four dimensions we think of the mass $\mu$ as arising from
Kaluza-Klein momentum around an additional periodic coordinate with period
$\beta \rightarrow \infty$.  This means the four-dimensional partition
function is
\be
- \log Z_{4d} = \beta \int_{-\infty}^\infty {d\mu \over 2 \pi}
\left(- \log Z_{3d}\right)\,.\label{Z4d}
\ee

The interaction between holes is clearly due to the off-diagonal
entries ${\cal O}_{12}$, ${\cal O}_{21}$.  If the holes are widely separated
these off-diagonal entries will be small and we can treat them
perturbatively.  The zeroth order term in this perturbation series is
independent of the separation distance, and the first order term vanishes identically.
So the leading contribution to the interaction
energy is 
\be
E_{\rm int} = - {1 \over 2\pi} \int_0^\infty d\mu \,
{\rm Tr} \, \left[({\cal O}_{11})^{-1} {\cal O}_{12} ({\cal O}_{22})^{-1} {\cal O}_{21}\right]\,.
\label{2a}
\ee
In this equation, for the inverses, one first constructs the diagonal projected 
operators ${\cal O}_{ii}$,
then inverts them in the subspace corresponding to hole $i$.
Writing the trace out in position space we get
\begin{widetext}
\be
E_{\rm int}  =  - {1 \over 2 \pi} \int_0^\infty d\mu \,
\int d^2x_1 d^2x_1' d^2x_2 d^2x_2' \,
\langle {\bf x}_1 \vert \left({\cal O}_{11}\right)^{-1} \vert {\bf x}_1' \rangle \,
\langle {\bf x}_1' \vert {\cal O}_{12} \vert {\bf x}_2 \rangle \,
\langle {\bf x}_2 \vert \left({\cal O}_{22}\right)^{-1} \vert {\bf x}_2' \rangle\,
\langle {\bf x}_2' \vert {\cal O}_{21} \vert {\bf x}_1\rangle
\label{Eint1}
\ee
\end{widetext}
where ${\bf x}_1,\,{\bf x}_1'$ correspond to points in the first hole and ${\bf x}_2,\,{\bf x}_2'$
to points in the second.

For widely-separated holes there are further simplifications we can make.  Large separation
necessarily means the separation distance $r$ is large compared to the size of the holes, so
we can pull the off-diagonal matrix elements out of the integrals.  Indeed we have (the projection
operators are immaterial for this)
\bea
\langle {\bf x}_1 \vert {\cal O}_{12} \vert {\bf x}_2 \rangle & \approx &
\langle 0 \vert \sqrt{-\nabla^2 + \mu^2} \, \vert r \rangle \nonumber\\
& = & - {1 \over 2 \sqrt{\pi}} \int_0^\infty {ds \over s^{3/2}} \, \langle 0 \vert e^{-s(-\nabla^2 + \mu^2)} \vert r \rangle \nonumber\\
& = & - {1 \over 2 \pi r^3} (1 + \mu r) e^{-\mu r}\label{prop}
\eea
where we introduced an integral representation for the square root and used the heat
kernel
\be
\label{HeatKernel}
\langle 0 \vert e^{s \nabla^2} \vert r \rangle = {1 \over 4 \pi s} e^{-r^2/4s}\,.
\ee
In principle the remaining matrix elements $\langle {\bf x}_1 \vert \left({\cal O}_{11}\right)^{-1} \vert {\bf x}_1' \rangle$, $\langle {\bf x}_2 \vert \left({\cal O}_{22}\right)^{-1} \vert {\bf x}_2' \rangle$
depend on $\mu$.  But for small holes we can neglect this dependence.  (Retaining it would
give corrections down by powers of $(\hbox{\rm size of hole})/r$.)  So using (\ref{prop})
in (\ref{Eint1}) and integrating over the explicit $\mu$ dependence gives the leading long-distance
interaction between small holes,
\be
E_{\rm int} = - {5 \, Q_1 Q_2 \over 32 \pi^3 r^7}
\label{potential}
\ee
where the charge associated with each hole is
\be
\label{charge}
Q_i = \int_{\hbox{\small hole $i$}} d^2x d^2x' \, \langle {\bf x} \vert \left({\cal O}_{ii}\right)^{-1} \vert {\bf x}' \rangle\,.
\ee
Note that $Q_i$ is simply the matrix element of $(P_i \sqrt{-\nabla^2} \, P_i)^{-1}$ between functions
that are constant (equal to one) in the hole.

\subsection{Round holes\label{round}}

The charge $Q$ defined in (\ref{charge}) depends on the geometry of the hole in question
but is independent of the separation distance.  Here we
study it for a round hole of radius $R$.

The approach we will use is quite simple: we adopt a lattice discretization of the operator
$P \sqrt{-\nabla^2} \, P$ and compute the relevant matrix element numerically.  In doing this we
can restrict to rotationally-invariant functions, meaning we only need to keep track of the radial
dependence.  Introducing a radial lattice spacing $a$ we identify
\bea
&& r \quad \leftrightarrow \quad \hat{r} = a
\left(\begin{array}{cccc}
1/2 & & & \\
& 3/2 & & \\
& & 5/2 & \\
& & & \ddots
\end{array}\right) \label{lattice1a}\\
&& {d \over dr} \quad \leftrightarrow \quad {1 \over 2a}
\left(\begin{array}{cccc}
0 & 1 & & \\
-1 & 0 & 1 & \\
& -1 & 0 & 1 \\
& & & \ddots
\end{array}\right)\label{lattice1b}
\eea
(note that we begin the lattice half-a-spacing from the origin, and that our discretization
of $d_r$ preserves antisymmetry).  The projection operator
\be
P \quad \leftrightarrow \quad \left(\begin{array}{cc}
\identity_{N \times N} & 0 \\
0 & 0
\end{array}\right)
\ee
where the rank of $P$ is related to the size of the hole by $R = a N$.  Finally
we identify the inner product on rotationally-invariant functions
\be
\int d^2x \, \phi_1^* \phi_2 \quad \leftrightarrow \quad \phi_1^\dagger \, \hat{r} \, \phi_2
\label{norm}
\ee
and the state of interest
\be
\label{state}
\int d^2x \, \vert {\bf x} \rangle \quad \leftrightarrow \quad \sqrt{2 \pi a}
\left(\begin{array}{l}
\left.\begin{array}{c} 1 \\ \vdots \\ 1 \end{array}\right\rbrace N \\
\left.\begin{array}{c} 0 \\ \vdots \end{array} \right.
\end{array}\right)
\ee
(the coefficient is fixed by requiring that the two states have the same norm).
From these ingredients we can construct $P \sqrt{- {1 \over r} d_r r d_r} \, P$ numerically as an
$N \times N$ matrix, invert it, and take the matrix element between the state
(\ref{state}).  Extrapolating to the continuum limit $N \rightarrow \infty$, $a \rightarrow 0$ with $R$ fixed,
we find that the charge for a round hole is
\be
\label{CircleCharge}
Q = 1.28 \, R^3 \qquad\quad \hbox{\rm (round hole)}
\ee
Note that the charge has units of $({\rm length})^3$, as could have been guessed
on dimensional grounds.  But one should be careful -- if two holes merge
to form a single hole the charges should not be expected to be additive.

\subsection{Rectangular holes\label{rectangle}}

In order to gain insight into how the charge depends on the shape of the hole, we now consider a rectangular
hole of size $L_1 \times L_2$.  Rather than use a position-space lattice, we will work in momentum space
with a mode cutoff.

The charge we wish to compute is
\[
Q = \int d^2x d^2x' \, \langle {\bf x} \vert (P \sqrt{-\nabla^2} P)^{-1} \vert {\bf x}' \rangle\,.
\]
Inserting complete sets of functions which vanish outside the hole, namely
\be
\label{modes}
\langle {\bf x} \vert mn \rangle = \left\lbrace
\begin{array}{ll}
{2 \over \sqrt{L_1 L_2}} \sin \big({m \pi x_1 \over L_1}\big) \sin \big({n \pi x_2 \over L_2}\big) &\hskip .1in \hbox{\rm in the hole} \\
\,\,\, 0 &\hskip .1in \hbox{\rm otherwise}
\end{array}\right.
\ee
with $m,n=1,2,3,\ldots$ we have
\be
\label{RectangleCharge}
{Q \over 64 L_1 L_2}= \sum_{{\rm odd}\, m, n, m', n'}\,
{1 \over m \pi} {1 \over n \pi}
\langle mn \vert {\cal O}^{-1} \vert m'n' \rangle
{1 \over m' \pi} {1 \over n' \pi}\,.
\ee
Here ${\cal O} = P \sqrt{-\nabla^2} \, P$.  The matrix elements of ${\cal O}$ in this basis are (assuming odd $m$ and $n$)
\begin{widetext}
\bea
\langle mn \vert {\cal O} \vert m'n' \rangle
&=& \langle mn \vert \sqrt{-\nabla^2} \vert m'n' \rangle \nonumber\\
\label{operator2}
&=& 64 L_1 L_2 \int {d^2 k \over (2\pi)^2} \, \sqrt{k_1^2 + k_2^2} \, \cos^2(k_1 L_1 / 2) \cos^2(k_2 L_2 / 2) f_1(m) f_2(n) f_1(m') f_2 (n')
\eea
\end{widetext}
where
\be
f_1(m) = {m \pi \over k_1^2 L_1^2 - m^2 \pi^2}, \hskip .1in
f_2(n) = {n \pi \over k_2^2 L_2^2 - n^2 \pi^2}
\ee
In the first line of (\ref{operator2}), we used the fact that the projection operators do not matter, and in the second line we made use of a Fourier
integral representation of the basis functions (\ref{modes}).

To compute the charge we introduce mode cutoffs $m \leq 2M-1$, $n \leq 2N-1$.  Then we can construct the operator ${\cal O}$ numerically
as an $(MN) \times (MN)$ matrix, invert it, and evaluate the sums in (\ref{RectangleCharge}).  Extrapolating to the continuum limit $M,N \rightarrow \infty$
then gives the charge.
For instance, for a square hole with $L_1 = L_2 = L$, this procedure gives the charge
\be
\label{SquareCharge}
Q = 0.228 \, L^3 \qquad\quad \hbox{\rm (square hole)}
\ee
By comparison, the charge of a round hole (\ref{CircleCharge}) is $0.230 \, ({\rm area})^{3/2}$.
So a square hole has slightly less charge than a round hole of equal area.

Another simple case is a rectangle with $L_1 \gg L_2$.  In this limit the matrix elements (\ref{operator2}) go over to
\be
\langle mn \vert {\cal O} \vert m'n' \rangle = \delta_{mm'} \, \langle n \vert \sqrt{-{d^2 \over dx_2^2}} \, \vert n' \rangle
\ee
i.e.,~it reduces to a one-dimensional problem.  (The easiest way to see
this is to rescale $k_1 \rightarrow k_1 / L_1$, $k_2 \rightarrow k_2 / L_2$ and send $L_1 \rightarrow \infty$.)
Using this in (\ref{RectangleCharge}) we find that the charge for a long thin rectangle is
\be
Q = L_1 Q_{\rm slit} \qquad\quad \hbox{\rm (thin rectangle)}\label{charge-rectangle}
\ee
where $Q_{\rm slit}$ is the charge for a slit of width $L_2$.  This is something we study in section \ref{slits}.  Borrowing the result (\ref{ExactSlit}),
the charge for a long thin rectangle is
\be
\label{ThinCharge}
Q = 0.393 \, L_1 L_2^2 \qquad \hbox{\rm for $L_1 \gg L_2$}
\ee
Note that, for a given area, a long thin rectangle has less charge than a square.  This is consistent with the observation above, that
the value of $Q/({\rm area})^{3/2}$ is slightly smaller for a square hole than for a round hole.  It's tempting to speculate that
round holes have the largest charge for a given area.

\subsection{Square hole: Perturbation theory\label{square-PT}}
The charge $Q$ in (\ref{RectangleCharge}) can also be calculated using the perturbation method
developed in \cite{Kabat:2010nm} where ${\cal O}$ is split into two parts: a term diagonal in the basis (\ref{modes}) and a nondiagonal term. The nondiagonal part is then treated in a perturbation series.

The $k_1$-integration in (\ref{operator2}) can be done by moving the integration contour slightly above the real $k_1$ axis and writing $ 4\cos^2(k_1L_1/2) = e^{ik_1L_1} + e^{-ik_1L_1} +2$.
For each term in this decomposition, the integration contour can be completed in the upper or lower half-plane, appropriately. We then pick up pole contributions at $\pm (m\pi /L_1) + i \epsilon$,
$\pm (m'\pi /L_1) + i \epsilon$ and cut contributions from 
the square root term $\sqrt{k_1^2 + k_2^2}$. The pole contributions cancel out unless $m = m'$.
For the cut contributions it is simpler to use the integral representation
\be
\sqrt{k_1^2 + k_2^2} = {1\over \pi} \int_{-\infty}^\infty d\gamma ~{k_1^2 + k_2^2 \over k_1^2 +k_2^2 + \gamma^2}
\label{sqrt-rep}
\ee
and evaluate contributions at the imaginary poles $k_1 = \pm i \sqrt{k_2^2 + \gamma^2}$.

We follow the same procedure for the $k_2$-integration. For a square hole, with
$L_1 = L_2 = L$, we then find
\be
\la m n \vert {\cal O}\vert m' n'\ra = \la m n \vert {\cal O}_{\rm pole} + {\cal O}_{\rm cut} \vert m' n'\ra
\ee
where
\be
\la m n \vert {\cal O}_{\rm pole} \vert m' n' \ra=
{1\over L} \sqrt{(m \pi )^2 + (n \pi )^2}~\delta_{mm'} \delta_{nn'}
\label{calO-square1}
\ee
is the diagonal term arising from the real-pole contributions in $k_1$ and $k_2$ integrations,
and
\begin{widetext}
\bea
\la m n \vert {\cal O}_{\rm cut}  \vert m' n'  \ra&=&
\!\!\!- {4\over \pi L} (m\pi )^2 \int_1^\infty \!\!\!\! dy~\sqrt{y^2 -1}\,(1+ e^{-m\pi y})
{n \pi \over (m\pi y)^2 +n^2 \pi^2} {n' \pi \over (m\pi y)^2 +n'^2 \pi^2}\,\, \delta_{mm'}\nonumber\\
&&\!\!\!- {16 \over \pi^2 L} \int_1^\infty \!\!\!\! dy~ \int_0^\infty\!\!\!\! dk \,k^2 \sqrt{y^2 -1}\, (1 + e^{-ky})
(1+ \cos k) {n \pi \over k^2 - n^2\pi^2} {n' \pi \over k^2 - n'^2\pi^2} {m \pi \over
k^2 y^2 + m^2 \pi^2}  {m' \pi \over k^2 y^2 + m'^2 \pi^2}\nonumber\\
 \label{calO-square2}
\eea
\end{widetext}
Treating ${\cal O}_{\rm cut}$ as a perturbation we have
\be
{\cal O}^{-1} =  {\cal O}_{\rm pole}^{-1} - {\cal O}_{\rm pole}^{-1} O_{\rm cut} {\cal O}_{\rm pole}^{-1} + \cdots\label{PT1}
\ee
with a corresponding series for the charge,
\[
Q = Q^{(0)} + Q^{(1)} + \cdots\,.
\]
For a square hole, denoting
\[
h(m,n) = {1 / \sqrt{(m\pi )^2 + (n \pi)^2}}\,,
\]
we find
\bea
Q^{(0)} &=& 64 L^3 \sum_{{\rm odd}\,m ,n} { h(m,n)\over (m\pi)^2 (n\pi)^2 }
= 0.170~L^3\nonumber\\
Q^{(1)} &=& -64 L^3 \!\!\!\!\!\!\sum_{{\rm odd}\,m,n,m',n'}\!\!\!\!\!\!\!{  \la m n \vert {\cal O}_{\rm cut} \vert  m' n' \ra  h(m,n) h(m',n')
\over (m\pi) (n\pi) (m'\pi) (n'\pi)}\nonumber\\
&=& 0.039 ~L^3 \label{PT2}
\eea
These first two terms capture, respectively, 75\% and 17\% of the lattice result (\ref{SquareCharge}), 
strongly suggesting the validity of the perturbation expansion.

\section{Interaction between slits\label{slits}}

\begin{figure}
\begin{center}
\scalebox{.75}{\includegraphics{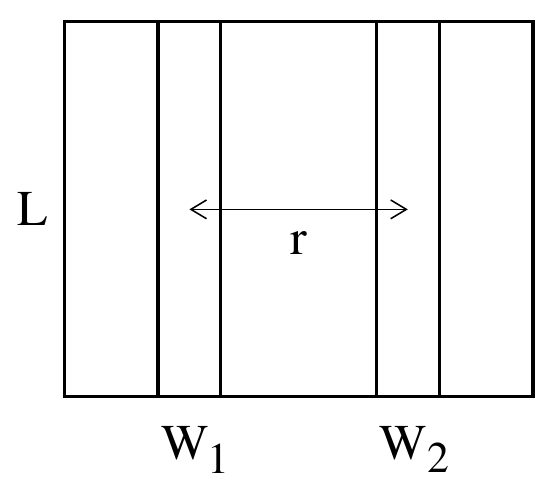}}
\end{center}
\caption{Slits of width $W_1$ and $W_2$ and length $L$, separated by a distance $r$.\label{slit}}
\end{figure}

As a simple setting which nicely illustrates our formalism, we conclude by studying a
plate with two long thin slits in it.  We denote the widths of the slits by $W$ and the
separation between slits by $r$.  We periodically identify both Euclidean time
with period $\beta \rightarrow \infty$ and the dimension along the slit with period $L
\rightarrow \infty$.  The geometry is shown in Fig.\ \ref{slit}.

The basic result for the interaction energy (\ref{Eint1}) still applies, with the
following modifications.
\begin{enumerate}
\item
The integral over Kaluza-Klein momentum becomes two dimensional,
\[
\beta L \int {d^2\mu \over (2 \pi)^2} = \beta L \int_0^\infty {\mu d\mu \over 2 \pi}
\]
\item
The operator ${\cal O}$ involves a one-dimensional Laplacian, and
the matrix element (\ref{prop}) changes to\footnote{For $d$ dimensional plates the heat kernel
(\ref{HeatKernel}) becomes $(4 \pi s)^{-d/2} \exp(-r^2/4s)$ and the matrix element (\ref{prop})
becomes $-2\left({\mu \over 2\pi r}\right)^{d + 1 \over 2} K_{(d+1)/2}(\mu r)$.}
\[
\langle x \vert \sqrt{-d_x^2 + \mu^2} \vert x' \rangle = - {\mu \over \pi r} K_1(\mu r)
\]
\end{enumerate}
This means that the interaction energy per unit length between narrow, widely-separated slits is
\bea
{E_{\rm int} \over L} & = & - {1 \over 4 \pi} \int_0^\infty \mu d\mu \,
\left({\mu \over \pi r} K_1(\mu r)\right)^2 Q_1 Q_2 \nonumber\\
& = & - 5.375 \times 10^{-3} \, {Q_1 Q_2 \over r^6}\label{Slit-energy}
\eea
where the charge associated with each slit is
\be
\label{SlitCharge}
Q = \int dx dx' \, \langle x \vert \big(P \sqrt{-d_x^2} \, P\big)^{-1} \vert x' \rangle\,.
\ee

\subsection{Lattice approach}

The charge (\ref{SlitCharge}) can be computed in a variety of ways.  We begin
with a lattice approach, in which we introduce a lattice spacing $a$ and
discretize the operators via
\bea
&& -{d^2 \over dx^2} \quad \leftrightarrow \quad {1 \over a^2}
\left(\begin{array}{cccc}
2 & -1 & & \\
-1 & 2 & -1 & \\
& -1 & 2 & -1 \\
& & & \ddots
\end{array}\right) \\
&& P = \left(
\begin{array}{ccc}
0 & & \\
& \identity_{N \times N} & \\
& & 0
\end{array}
\right)
\eea
The rank of $P$ is related to the width of the slit by $W = aN$.  Finally we identify
\be
\label{state2}
\int dx \, \vert x \rangle \quad \leftrightarrow \quad \sqrt{a}
\left(\begin{array}{l}
\, 0 \\ \, 1_N \\ \, 0
\end{array}\right)
\ee
Computing the matrix element and extrapolating to the continuum limit gives the
charge for a slit,
\be
\label{ExactSlit}
Q = 0.393 \, W^2\,.
\ee
The same coefficient appears in the charge of a long thin rectangle (\ref{ThinCharge}).

\subsection{Perturbation theory}

An alternative approach to computing the charge of a slit is to use
perturbation theory.  Inserting a complete set of states
\be
\langle x \vert n \rangle = \sqrt{2 \over W} \, \sin \big ({n \pi x \over W}\big) \qquad n = 1,2,3,\ldots
\label{slit-modes}
\ee
the charge (\ref{SlitCharge}) becomes
\be
Q = 8 W \sum_{\hbox{\small $m,n$ odd}} {1 \over m \pi} {\cal O}^{-1}_{mn} {1 \over n \pi}
\label{Slit-charge2}
\ee
where ${\cal O} = P \sqrt{-d_x^2} P$.  The matrix elements of this operator were studied in \cite{Kabat:2010nm},
where it was found that they could be decomposed into `pole' and `cut' contributions,
${\cal O} = {\cal O}^{\rm pole} + {\cal O}^{\rm cut}$.  The matrix elements for odd $m$ and $n$ are\footnote{These
are the even-parity modes of \cite{Kabat:2010nm} equations (29), (32) with the dictionary $p = m/2$, $q = n/2$, $a = W/2$.}
\bea
&& {\cal O}^{\rm pole}_{mn} = {m \pi \over W} \, \delta_{mn}\label{calO-slit}\\
&& {\cal O}^{\rm cut}_{mn} = - {4 \over \pi W} \int_0^\infty dy \, y \big(1 + e^{-y}\big)
{m \pi \over m^2 \pi^2 + y^2} \, {n \pi \over n^2 \pi^2 + y^2}\nonumber
\eea
The pole contribution corresponds to putting a Dirichlet boundary condition on the edges of the slit, while
the cut contribution captures diffractive effects.  Once again, treating ${\cal O}^{\rm cut}$ as a perturbation as in (\ref{PT1}), 
the first few terms in the series for the charge are
\bea
&& Q^{(0)} = {7 \zeta(3) \over \pi^3} \, W^2 = 0.2714 \, W^2 \nonumber\\
&& Q^{(1)} = 0.0773 \, W^2 \label{Slit-Qs}\\
&& Q^{(2)} = 0.0268 \, W^2\nonumber
\eea
All terms in this perturbation series are positive.  Note that the perturbation series appears to be nicely convergent.  Compared to the lattice result
(\ref{ExactSlit}), the first three terms capture 69\%, 20\% and 7\% of the exact result, respectively, for a
total of 96\%.

\section{Universality of $1/r^{7}$}\label{universality}

The behavior of the potential between two holes in a plate is very reminiscent of the van der Waals interaction between neutral atoms of zero intrinsic dipole moment. In that case, there is an argument due to Feinberg and Sucher \cite{FS} which shows that the behavior of the potential is universal at large distances.
In modern language the argument for this low energy theorem is that the effective action 
is of the form ${\half} ( \alpha_E E^2 + \alpha_B B^2 )$, where $E$ and $B$ are the electric and magnetic fields, respectively, and $\alpha_E, \alpha_B$ are the corresponding polarizabilities.
If one introduces a field $\Phi$ representing the creation and annihilation of the atom as a whole,
then the effective action is of the form
\be
S_{\rm eff} = g_1 \del_\mu \Phi \del _\nu \Phi  F^{\mu \alpha} F^\nu_\alpha
~+~ g_2 \Phi^2 F^2 ~+\cdots \label{u1}
\ee
For simplicity, we take the atom to be represented by a scalar field, the final result is not sensitive to this choice. The terms displayed in (\ref{u1}) are those with the  lowest dimension; these are the relevant ones for low energy or long distance behavior. 
(The coefficients $g_1$ and $g_2$ can be related to the polarizabilities, but this is not relevant for our argument.)
This effective action can be used to calculate the potential between two atoms at large separation;
it is given by the two-photon exchange generated by (\ref{u1}).
The computation of this process leads to the $1/r^7$ behavior of the van der Waals potential, in the long-distance regime
where retardation effects are important \cite{FS}. This is the universality of the van der Waals interaction.

A similar low energy theorem applies to the potential between two holes on a plate.
For this purpose, we first note that the boundary action (the action on the holes) can be obtained from the quantum effective action $\Gamma [\chi ]$ which generates the 1PI diagrams. For this, recall that the boundary action is defined by
\be
\exp \left( - S [\phi_0 ] \right) =
\int [d \phi] ~\exp \left( - S [ \chi + \phi ]\right)
\label{u2}
\ee
where $\chi$ obeys the condition $ \chi \rightarrow \phi_0 $ as one approaches the boundary.
We defined $\chi$ as a special solution of the (free) equations of motion with this boundary condition. Explicitly, for fields on the left side of the partition of the box in which the theory is defined, we could take
\be
\chi (x) = \int d^{d-1}x' ~\phi_0 (x') \,n \cdot \del' G(x\vert x') 
\label{u3}
\ee
where $G (x\vert x')$ is the propagator to the left (and right) of the plate with Dirichlet boundary conditions.
We do not need to make this particular choice, $\chi$ is any specific field with the boundary behavior $\chi \rightarrow \phi_0$, so that $\phi$ in (\ref{u2}) can be taken to obey Dirichlet boundary conditions.

Consider now the calculation of $\Gamma [\chi ]$ for the left side of the partition of the theory in a box.
From the Legendre transformation of the generating functional for connected Green's functions, it is easily seen that we can write
\be
\exp \left( - \Gamma [\chi ]  \right)
= \int [d\phi ] ~\exp \left( - S[\chi + \phi ] + \int {\delta \Gamma \over \delta \chi} \phi
\right)
\label{u4}
\ee
where $\chi$ is, for the moment, an arbitrary field. If we choose $\chi$ to be a solution of the
quantum equations of motion, $\delta \Gamma /\delta \chi ~=0$, obeying the condition
$\chi \rightarrow \phi_0$ on the boundary, we see that the functional integral on the right side of
(\ref{u4}) becomes the defining integral (\ref{u2}) for the boundary action. In other words,
\be
S [\phi_0 ] = \Gamma [ \chi ],  \hskip .2in {\delta \Gamma \over \delta \chi } =0,
\hskip .2in \chi \rightarrow \phi_0 ~~{\rm on~ the~ boundary}
\label{u5}
\ee
This gives the boundary action for any theory, including the effect of interactions. It is the quantum effective action evaluated on its critical point with the boundary condition
$\chi \rightarrow \phi_0$.

The long-distance interaction between holes comes from terms in the effective action which are quadratic in
$\chi$.  For these quadratic terms, the low energy or derivative expansion of $\Gamma [\chi ]$ follows the
familiar pattern. For a scalar field, the lowest order term
is the mass term. Since we are considering a massless theory, we can take the renormalized mass to be zero, so there is no $\chi^2$ term in $\Gamma [\chi ]$. The lowest nontrivial term in $\Gamma [\chi ]$ is then
the kinetic term,
\be
\Gamma [\chi ] = {1\over 2} \int \del_\mu  \chi \, \del^\mu \chi ~+\cdots
\label{u6}
\ee
This term has been canonically normalized by an appropriate wavefunction renormalization.
It leads to the boundary action we used above and gives $1/r^7$ behavior for the hole-hole potential on the plate. Higher dimension terms in $\Gamma [\chi ]$ will not contribute to the long distance behavior of the potential. This argument makes it evident that we will get the same behavior for other kinds of fields as well.
For example, for the electromagnetic field, $\Gamma [A ]$ has similar behavior,
with the transverse potentials contributing. Hence it will lead to the same 
$1/r^{7}$
potential between holes.  It would be interesting to extend this result
to spinor fields.

Thus we have obtained a low energy theorem, analogous to the Feinberg-Sucher proof of universality of the van der Waals force: The long distance potential
between two holes in a plate behaves as $1/r^7$ for all massless fields for
which the leading kinetic operator is the Laplacian. (This result holds for a four-dimensional theory.)

\section{Discussion\label{discussion}}

In this paper we have analyzed the Casimir interaction between two holes on a conducting plate
(a plate on which the fields obey Dirichlet conditions) using the formalism of the non-local field theory developed in \cite{Kabat:2010nm}. For a separation distance $r$ which is large compared to the linear dimensions of the holes, the interaction energy is proportional to $Q_1 Q_2/r^{7}$, where $Q_1, Q_2$ are charges associated with the holes. The $1/r^{7}$ form is universal, given by a low energy theorem. For any number of holes, from equation (\ref{2a}), the result is the same with a pairwise interaction between holes. (For long slits, because of an additional integration along the length of the slits, the interaction energy scales as
$1/r^{6}$.)

There are clearly interesting questions for further analysis. The charge $Q$ for each hole depends on the geometry of the hole. While we have considered round and rectangular holes and long slits explicitly, it is
interesting to seek a more general understanding of how $Q$ is related to the geometry of the hole. 

The $1/r^{7}$ behavior of the two-hole interaction is for large distances. What is the behavior of the interaction energy as the holes come closer? While we do not have a definitive answer, it is worth noting
that we are considering the mutual interaction energy between holes; the self-energies of the holes do not
appear, or, equivalently, they have been subtracted out. But when the holes merge, the entire energy is the
self-energy. It would be interesting to study the approach and merger of two holes in more detail.  An
intriguing possibility is that, as the holes come closer, the potential could reach a minimum value at some
finite separation, similar to the Lennard-Jones potential between neutral atoms.
 
A large number of small mobile holes on a plate would behave like a gas of
particles with a pairwise (two-particle) interaction which is attractive and
goes like $1/r^{7}$ at long separations.  It would be interesting to study the
thermodynamics of this ``Casimir gas''.

Our analysis was done using a non-local field theory on the holes. There are other methods of analysis, notably the world-line formalism, which has been used before for studying diffractive effects on Casimir forces \cite{gies}. It would be interesting, and an independent check, to reproduce our results in this formalism.

Finally, one may ask about experimental observation of this interaction. Casimir energies resulting directly from fundamental forces, such as the electrodynamic Casimir energy,  are small and very difficult to measure \cite{lamoreaux}. However the thermodynamic analogue of the Casimir effect, which can occur in liquid mixtures near a critical point, can be appreciable because, for this effect,  temperature plays the role of
Planck's constant. In fact, there have been recent observations of such a thermodynamic analogue of the Casimir effect \cite{colloids}. It could be that the interaction between holes would be observable in such  a setting.

\bigskip
\centerline{\bf Acknowledgements}
We are grateful to Alexios Polychronakos for valuable discussions.
This work was supported by U.S.\ National Science Foundation grants
PHY-0855582, PHY-0758008 and PHY-0855515 and by PSC-CUNY awards.


\begin{thebibliography}{99}

\bibitem{Casimir:1948dh} H.B.G. Casimir, Proc. K. Ned. Akad. Wet {\bf 51}, 793
(1948); Indag. Math. {\bf 10}, 261 (1948); H.B.G. Casimir
and D. Polder, \PR~{\bf 73}, 360 (1948).

\bibitem{Milton:2008st} K.A. Milton, {\it Recent developments in Casimir effect},
J. Phys. Conf. Ser. {\bf 161}, 012001 (2009) [arXiv:0809.2564];
K. A. Milton, {\it The Casimir Effect: Physical Manifestations of Zero-Point Energy} (World Scientific, 2001); M. Bordag, U. Mohideen and V.M. Mostepanenko, Phys. Rept. {\bf 353}, 1 (2001).

\bibitem{Kabat:2010nm}  D. Kabat, D. Karabali and V.P. Nair,
{\it Edges and diffractive effects in Casimir energies}, arXiv:1002.3575[hep-th].

\bibitem{Jackiw:1988sf} R. Jackiw,
{\it Analysis on infinite dimensional manifolds: Schroedinger representation for quantized fields}.
In {\it Field theory and particle physics: proceedings of the 5th Jorge Andre Swieca Summer School}, 
O.\ Eboli, ed.\ (World Scientific, 1990).

\bibitem{Hatfield:1992rz} B. Hatfield, {\it Quantum field theory of point particles and strings}
(Addison-Wesley, 1992).  See section 10.1.

\bibitem{gies} H. Gies and K. Klingm\"uller, \PRL~{\bf 97}, 220405 (2006);
J. Phys. {\bf A39}, 6415 (2006).


\bibitem{MIT2} N. Graham {\it et al}, arXiv:0910.4649 [quant-ph].

\bibitem{FS}  G. Feinberg and J. Sucher, Phys. Rev. {\bf A2}, 2395 (1970);
C. Itzykson and J-B. Zuber, {\it Quantum Field Theory} (McGraw-Hill, 1980), p. 365.

\bibitem{lamoreaux} See, for example, the articles in \cite{Milton:2008st}, and also S.K. Lamoreaux, Rep. Prog. Phys. {\bf 68}, 201 (2005)

\bibitem{colloids} C. Hertlein {\it et al}, Nature {\bf 451}, 172 (2008); F. Soyka {\it et al}, \PRL~{\bf 101}, 208301 (2008).

\end{thebibliography}
\end{document}